# Faster Fair Solution for the Reader-Writer Problem

*Vlad Popov and Oleg Mazonka*
2013

**Abstract** *A fast fair solution for Reader-Writer Problem is presented.*

## 1. Introduction

A classical Reader-Writer Problem is a situation where a data structure can be read and modified simultaneously by concurrent threads. Only one Writer is allowed access the critical area at any moment in time. When no Writer is active any number of Readers can access the critical area. To allow concurrent threads mutually exclusive access to some critical data structure, a mutually exclusive object, or mutex, is used. As an implementation mutex is a special case of a more generic synchronisation concept – semaphore. A simple image of a semaphore is a positive number which allows increment by one and decrement by one operations. If a decrement operation is invoked by a thread on a semaphore whose value is zero, the thread blocks until another thread increments the semaphore.

## 2. Background

Below we denote a semaphore by Semaphore(*n*), where *n* is an initialising number. Decrement and increment operations we denote as Wait and Signal correspondingly. In this notation a mutex is Semaphore(1), so we use only semaphore concept from now on.

The immediate and straightforward solution to Reader-Writer Problem [1] involves setting up a semaphore with a counter initialised to a number of Readers allowed simultaneously read the critical section. A Writer then sequentially reduces the semaphore counter by that number by waiting until all Readers finish and at the same time not allowing new Readers to start. This solution has too many deficiencies: assumption that the semaphore is buffered, necessity to know upfront a number of possible Readers, and a linear loop depending on the number of Readers.

Contrary to the above, a classical solution presented below is simple and efficient.

| Initialisation | Reader | Writer |
|---|---|---|
| mx = Semaphore(1) <br> wrt = Semaphore(1) <br> ctr = Integer(0) | - Wait mx <br> - if (++ctr)==1, then Wait wrt <br> - Signal mx <br><br> [Critical section] <br><br> - Wait mx <br> - if (--ctr)==0, then Signal wrt <br> - Signal mx | - Wait wrt <br><br> [Critical section] <br><br> - Signal wrt |

The only downside it has is the starvation of the Writer: a Writer thread does not have a chance to execute while any number of Readers continuously entering and leaving the working area.

To avoid this problem the following commonly known solution is proposed.

```
Initialisation        Reader                              Writer

in = Semaphore(1)     - Wait in                           - Wait in
mx = Semaphore(1)     - Wait mx                           - Wait wrt
wrt = Semaphore(1)    - if (++ctr)==1, then Wait wrt
ctr = Integer(0)      - Signal mx                         [Critical section]
                      - Signal in
                                                          - Signal wrt
                      [Critical section]                  - Signal in

                      - Wait mx
                      - if (--ctr)==0, then Signal wrt
                      - Signal mx
```

This solution is simple and fast enough. However the penalty in comparison to the previous one is that the Reader must lock two mutexes to enter the working area. If the working area is fast and assuming that mutex locking is a heavy system call, there would be a benefit of having an algorithm which allows locking one mutex on entering the working area and one on exiting.

## 3.   Our solution

The following presented solution alleviates the simultaneous use of two mutexes for a Reader.

```
Initialisation         Reader                             Writer

in = Semaphore(1)      - Wait in                          - Wait in
out = Semaphore(1)     - ctrin++                          - Wait out
wrt = Semaphore(0)     - Signal in                        - if (ctrin==ctrout)
ctrin = Integer(0)                                          then Signal out
ctrout = Integer(0)    [Critical section]                   else
wait = Boolean(0)                                             - wait=1
                       - Wait out                             - Signal out
                       - ctrout++                             - Wait wrt
                       - if (wait==1 && ctrin==ctrout)        - wait=0
                         then Signal wrt
                       - Signal out                        [Critical section]

                                                          - Signal in
```

The main idea here is that a Writer indicates to Readers its necessity to access the working area. At the same time no new Readers can start working. Every Reader leaving the working area checks if there is a Writer waiting and the last leaving Reader signals Writer that it is safe to proceed now. Upon completing access to the working area Writer signals waiting Readers that it finished allowing them to access the working area again.

## 4.   Conclusion

In this paper we presented a faster algorithm for Reader-Writer Problem. Fair algorithm works without starvation for buffered (FIFO) semaphores. It also works with no starvation with semaphores where waiting threads selected randomly by the operating system, because the probability that thread does not get a chance to execute decreases with time exponentially. In case of FILO semaphores Morris algorithm [2,3] can be used to avoid starvation.

Our solution requires only one mutex locking for a Reader both entering the critical section and exiting the critical section; and Writer needs two mutex lockings for Writer to enter critical section, and no locking for exiting. Usage of extra mutex, as in traditional solution, may not be significant in most cases. In some cases, however, when the critical data is accessible in fast manner the benefit of our solution may be substantial.